# One-dimensional Weak Localization of Electrons in a Single InAs Nanowire

*Dong Liang[§], Mohammed R. Sakr[†§] and Xuan P.A. Gao[*]*

Department of Physics, Case Western Reserve University, Cleveland, OH 44106.

[§]These authors contributed equally to this work. [†]Permanent address: Department of physics, Faculty of Science, Alexandria University, EGYPT. [*]Corresponding author. Email: xuan.gao@case.edu

**ABSTRACT**

We report on low temperature (2-30K) electron transport and magneto-transport measurements of a chemically synthesized InAs nanowire. Both the temperature, *T*, and transverse magnetic field dependences of the nanowire conductance are consistent with the functional forms predicted in one-dimensional (1D) weak localization theory. By fitting the magneto-conductance data to theory, the phase coherence length of electrons is determined to be tens of nanometers with a $T^{-1/3}$ dependence. Moreover, as the electron density is increased by a gate voltage, the magneto-conductance shows a possible signature of suppression of weak localization in multiple 1D subbands.

**MANUSCRIPT**

The downscaling of electronics, achieved during the last half century, faces increasing technological and fundamental challenges. Semiconductor nanowires (NWs) offer a realistic addition to the semiconductor industry since they can be prepared with reproducible and controllable electronic properties in high-yield, as required for large-scale integrated nanoelectronic systems [1-3]. In addition, similar to carbon nanotubes [4], gated semiconductor NWs represent a unique quasi-one dimensional (1D) system for exploring physical phenomena at the nanoscale, yet with many variations of material choices. In particular, InAs nanowires [5] are promising because of their high electron mobility and strong quantum confinement effects. Besides, InAs nanowire field effect transistors (FETs) are shown to have at least comparable noise performance as that of planar InAs high electron mobility transistors [6]. Recent demonstration of band-engineered InAs/InP radial nanowire heterostructures with high electron mobility [7] also generated further interests in InAs nanowire as a clean system for studying electron transport phenomena in 1D [8].

Constructive quantum interference tends to localize electrons as they diffuse coherently in weakly disordered systems. Such weak localization effect has been widely observed in various disordered electronic systems[9, 10]. Electrons have to maintain their phase coherence for quantum-interference effects to take place. Therefore, the weak localization diminishes at high temperatures where frequent electron-phonon and electron-electron scatterings destroy the coherence. Furthermore, a magnetic field introduces different phases for electron traveling in time-reversed paths and thus suppresses the weak localization and enhances the sample conductance. Because the weak localization correction to conductance is directly related to the phase coherence length $L_\varphi$, analyzing the weak field magneto-conductance is often used to extract $L_\varphi$, which is a key parameter for any device operation based on electron coherence in nanostructures.

Recently, low temperature magneto-transport measurements were used to study the electron localization in carbon nanotubes [11] and silicon NWs [12]. Magneto-transport was also investigated in arrays of GaN [13] and InAs NWs [14] to extract the phase coherence length and spin-orbit (SO) scattering length. Here, we report on low temperature magneto-conductance $G(B)$ measurements of a *single* gated InAs nanowire in transverse magnetic field. The observed positive magneto-conductance $G(B)$ data is consistent with the 1D weak localization effect and we extract a gate voltage controllable phase coherence length $L_\varphi$ in the range of 20-50nms in our temperature range (2-30K). The temperature dependence of the phase-coherence length $L_\varphi \sim T^{-1/3}$ is consistent with the one-dimensional Nyquist phase breaking mechanism [15, 16]. More interestingly, in our single nanowire device with gate tunable electron density $n$, we observe possible signature of suppression of weak localization associated with multiple 1D subbands as $n$ is increased.

Based on the vapor transport method reported earlier [7], InAs nanowires were grown on silicon/silicon dioxide ($Si/SiO_2$) substrate in a thermal chemical vapor deposition (CVD) system. Ultra-sound sonication was used to suspend the nanowires in isopropyl alcohol (IPA) solution which was dropped onto 290 nm thick oxide on a highly doped *n*-type Si substrate which was used as a gate electrode. Photolithography and electron beam evaporation of Ti/Al (2nm/60nm) were used to establish source and drain contacts that are 2 $\mu$ m apart. The sample was dipped in buffered hydrofluoric acid (HF) solution for about 3 s before evaporation to remove any native oxide and ensure Ohmic contacts. The two-terminal conductance ($G$) of the nanowire was measured by low frequency lock-in technique in a Quantum Design PPMS cryostat [17]. We have measured two InAs NWs with 20nm diameter and both devices yielded similar results. Here we present representative data from one device.

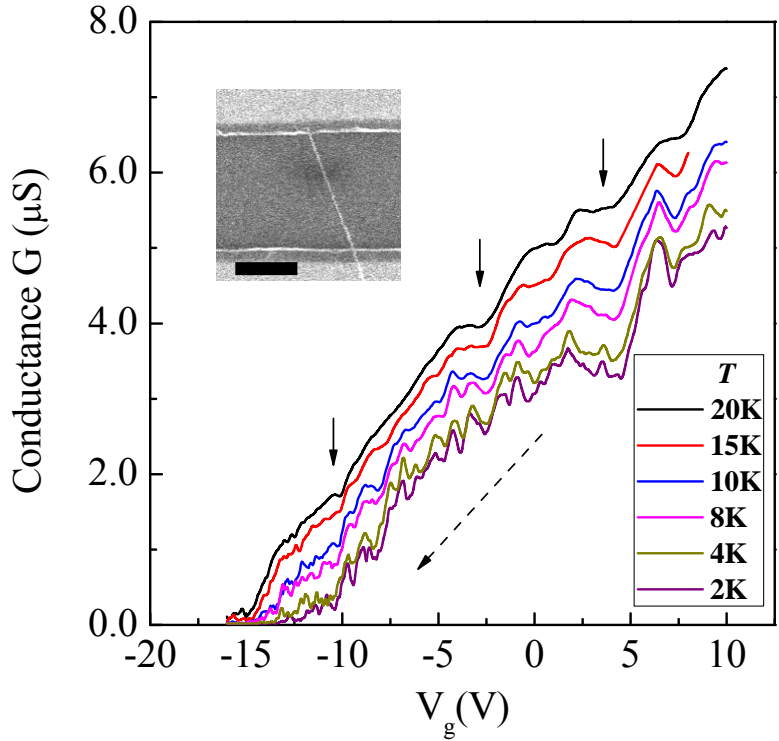

**Figure 1** Gate voltage ($V_g$) dependent conductance ($G$) of a 20nm InAs nanowire at temperatures of 20, 15, 10, 8, 4, and 2K. The dashed arrow shows the sweep direction of $V_g$. A few dominant dips in the $G(V_g)$ curves are marked by the vertical arrows, showing that the dip positions do not shift in $V_g$ while conductance decreases as $T$ reduces. The inset is the SEM image of the InAs NW device with channel length $L$=2.0 μm. Scale bar is 1μm.

The conductance of the InAs nanowire as a function of the gate voltage, $V_g$, is shown in Fig. 1 for different temperatures at zero magnetic field. The scanning electron microscope image of the nanowire is shown in the inset of Fig. 1. The source-drain voltage used in the measurement is 3 mV. The sample showed slight hysteresis effect in $G(V_g)$ with respect to the sweep direction of $V_g$. For clarity purpose we show the $G(V_g)$ data for one sweep direction of $V_g$, as indicated by the dashed arrow. There are two main features in this set of $T$-dependent $G(V_g)$ traces. First, there are multiple oscillations on top of a smooth background of $G(V_g)$. Some dominant oscillations with large $V_g$ spacing survived even at 20K, as marked by the solid arrows. These oscillations are likely to originate from the singularities in the density of states of the quantized 1D subbands, similar to the observations in silicon NWs with small diameter [18, 19]. Thus multiple subbands can contribute to the transport in our system depending on the gate voltage. Secondly, as temperature is lowered, device conductance decreases for all the $V_g$'s in this temperature range. There could be two implications for the decrease of conductance: it could be due to a decrease in either the electron density or mobility (or both). However, the positions of the dominant dips in $G(V_g)$ do not show any considerable displacement at different $T$'s. This suggests that the positions of different subband edges (in the axis of $V_g$) do not shift with $T$ and as result the electron density is constant for a given $V_g$. Therefore, we conclude that the temperature dependent conductance in our sample is not a trivial carrier density effect.

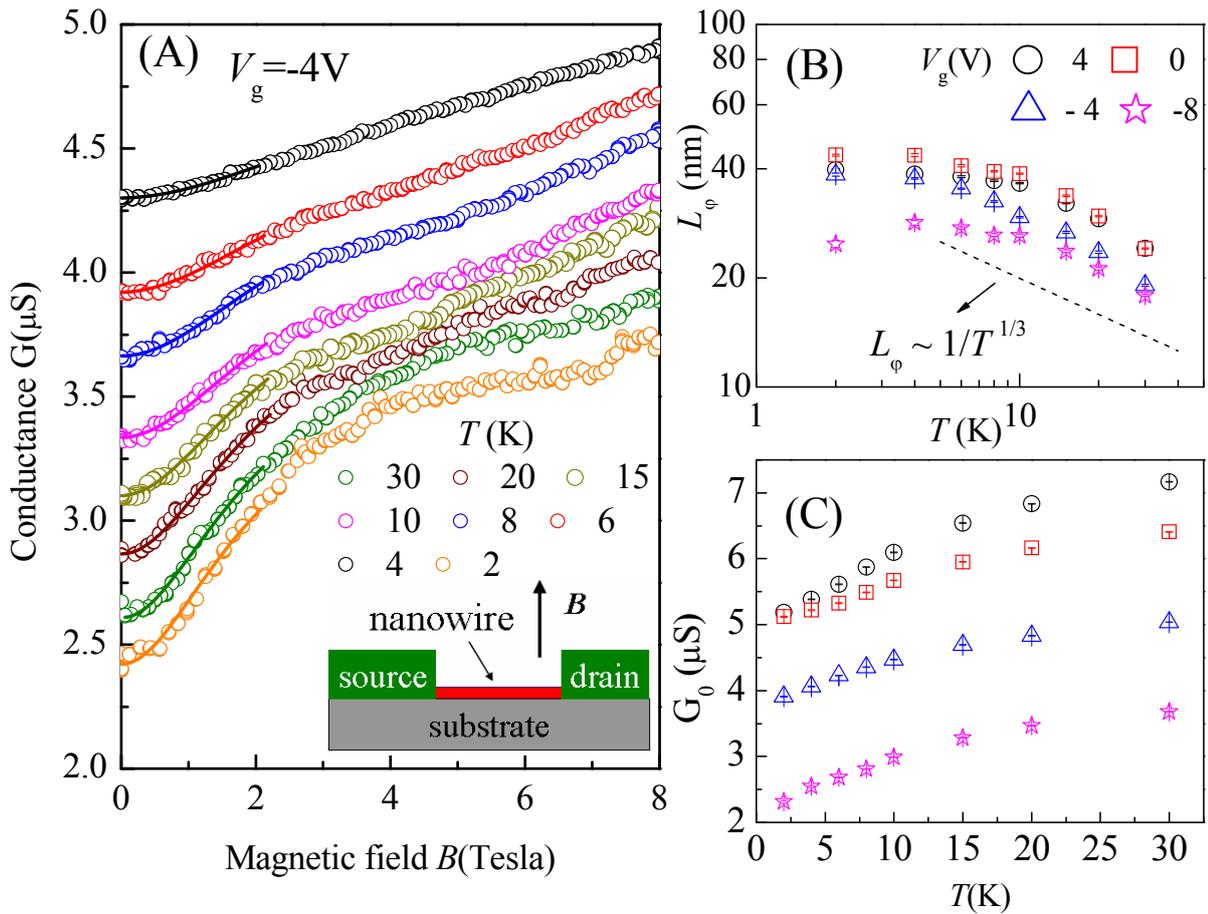

**Figure 2** (A) Magneto-conductance $G(B)$ at $V_g$ = -4V, and temperature 30, 20, 15, 10, 8, 6, 4, and 2K. Solid colored curves are fits to 1D weak localization theory. The inset shows the measurement configuration. (B) Electron phase coherence length extracted from the 1D weak localization fitting to the data at various $V_g$ and temperatures. (C) Residual Drude conductance vs. $T$, obtained from fitting $G(B)$ to 1D weak localization.

Figure 2(A) shows the dependence of the magneto-conductance, $G(B)$, on the magnetic field, $B$, for a number of temperatures at $V_g = -4$ V. The magnetic field was applied perpendicular to nanowire axis and the substrate, as shown in the inset panel of Fig. 2(A). It is clear that the magneto-conductance is positive and the amount of conductance increase in $B$ is larger at lower temperature. At low temperatures, $G(B)$ saturates around $B = 4$T while it increases more slowly and continuously at $T = 30$K. The positive magneto-conductance is due to the suppression of the weak localization by the magnetic field. In one dimension, the weak localization correction to the conduction is given by [15, 20]

$$G(B) = G_0 - \frac{2e^2}{hL}\left(\frac{1}{L^2_\varphi} + \frac{e^2 B^2 w^2}{3\hbar^2}\right)^{-1/2}. \qquad (1)$$

Here, $L = 2\mu$m is the length of the nanowire, $w = 20$nm is its width or diameter, and $G_0$ is the classical Drude conductance without the localization correction. There are only two parameters $L_\varphi$ and $G_0$ in Eq.1 used to fit to our experimental data at low magnetic field. The fitting curves are indicated by the solid lines in Fig. 2(A) and the fitted $L_\varphi$ are shown in Fig. 2(B) for different $V_g$'s. Note that Eq.1 is valid only for $L_\varphi > w$ which is satisfied from the fitting results in Fig.2(B). The extracted phase coherence lengths are seen to follow a power law dependence of temperature $L_\varphi \sim T^{-1/3}$, consistent with the dephasing mechanism being the electron-electron collisions with small energy transfers (or the 'Nyquist' dephasing) as observed previously in many other systems [11-16]. Notably, $L_\varphi$ tends to saturate at $T < 10$K in Fig. 2(B). Similar saturation phenomenon was attributed to either the overheating of electrons [12] or other intrinsic [21] or extrinsic [16e] dephasing mechanisms. It is interesting to note that unlike a previous report on arrays of InAs nanowires [14], the interpretation of our data does not need to involve weak-antilocalization effect which would be indicated as a negative magneto-conductance. This means that the spin-flipping length is much longer than $L_\varphi$ in our system, suggesting the potential use of InAs nanowires in applications requiring long spin coherence length. The fitted Drude conductance $G_0$ is plotted against temperature in Fig. 2(C) for all four $V_g$'s we studied in detail. Figure 2(C) shows that $G_0$ still has a weak insulating behavior after correcting for the weak localization, indicating that there is additional mechanism contributing to the temperature dependence of the sample conductance. We will return to the temperature dependence of $G$ later.

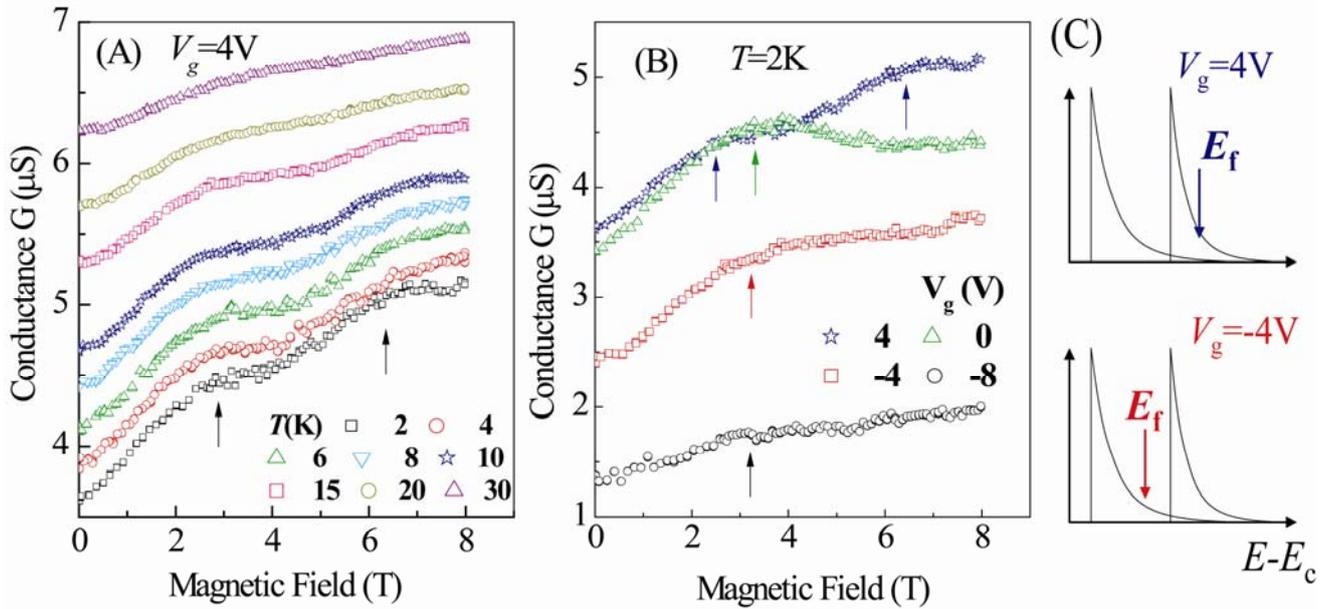

**Figure 3** (A) Magneto-conductance $G(B)$ at $V_g$= 4V, and $T$= 2, 4, 6, 8, 10, 15, 20, and 30K. For this gate voltage where electron density is relatively high, $G(B)$ exhibits two saturation fields marked by the black arrows. This is attributed to the existence of two 1D-subbands with very different mobilities and phase coherence lengths. (B) $G(B)$ at $T$=2K for four gate voltages. At low electron density ($V_g$=-8, -4V) there is only one saturation field for suppression of weak localization effect (marked by colored arrows). As the gate voltage is increased to $V_g$=4V, two saturation fields appear in the $G(B)$ curve. (C) Schematics of electron density of states in NW vs energy above the bottom of conduction band. The bottom (top) panel shows the occupation of one (two) subband(s) for low (high) $V_g$.

With the gate voltage, we can control the electron density $n$ and thus adjust the population of different quantized 1D subband in NWs. We observed that the magneto-conductance $G(B)$ showed increased complexity related to the occupation of different 1D subbands as the electron density was increased. We studied the temperature dependence of $G(B)$ at four $V_g$'s: -8, -4, 0 and 4V. $G(B)$ curves for the lower two electron densities ($V_g$=-8, -4V) look qualitatively similar to Fig.2 (A), i.e. there is a positive $G(B)$ which saturates at a few Tesla. As the gate voltage is increased to $V_g$=4V [22], two saturation fields appear in the $G(B)$ curve. The $G(B)$ traces for $V_g$=4V are shown in Fig. 3(A) for various temperatures, and the two saturation fields are marked in the two-step like $G(B)$ curves at low $T$. The weakening of both steps in $G(B)$ at elevated temperatures and the positive sign of magneto-conductance suggest that both are related to the suppression of weak localization effect. We also compare the $G(B)$ data for all four $V_g$'s at $T$=2K in Fig. 3(B) to show the evolution of $G(B)$ curve as electron density is increased.

We now discuss the meaning of the saturation fields in $G(B)$ in more detail. In the weak localization theory, the suppression of quantum interference becomes significant when the magnetic flux enclosed in the closed loop of electron diffusion path is on the order of flux quanta $h/2e$, where $h$ is the Planck's constant and $e$ is the electron charge. This characteristic magnetic field, associated with the suppression of weak localization, can be estimated to be $B_c \sim \frac{h}{2e}/w \times L_\varphi$ for nanowire with width $w<L_\varphi$. As $B$ increases beyond $B_c$, the weak localization will be completely destroyed above a magnetic field $B_s \sim \frac{h}{2e \times l^2}$ with $l$ being the electron mean free path [23]. If we use $w$=20nm and $L_\varphi$=40nm, we calculate $B_c$=2.6T. And similarly, a $B_s$=3T will yield an electron mean free path of ~26nm, corresponding to an overall sample conductance scale of $e^2/\hbar \times l/L \sim 3\mu S$, which is indeed a reasonable value for the typical conductance of our nanowire. Therefore, the typical saturation field of 3-4T we observe in $G(B)$ measurements is in good agreement with the interpretation of complete destruction of weak localization effect. The second saturation field for $V_g$=4V at higher magnetic field (~7Tesla) would be related to a subband with $L_\varphi$ and $l$ at nearly half of the values we mentioned above. The gradual filling of additional subband as $V_g$ increases from -8 to 4V is illustrated in Fig. 3(C).

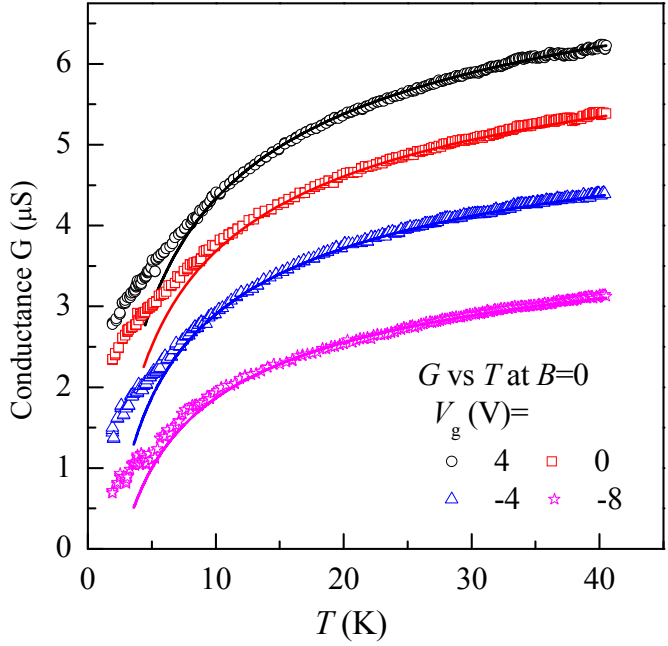

**Figure 4** Conductance vs temperature ($G(T)$) data for electrons in InAs nanowire with 20nm diameter at gate voltages of $V_g$= 4, 0, -4, and -8V. The magnetic field is zero. The solid lines are fits to the 1D weak localization with $L_\varphi \propto T^{-1/3}$. The deviation between data and fit at $T<10$K is due to the saturation of $L_\varphi$ as shown in Fig. 2(B).

We now discuss the temperature dependence of the conductance at zero magnetic fields which is shown in Fig. 4 for all four different gate voltages. $G(T)$ at $B=0$ is given by setting $B=0$ in Eq.1:

$$G(T) = G_0 - \frac{2e^2}{h}\frac{L_\varphi}{L} \qquad (2).$$

Since we have found $L_\varphi \propto T^{-1/3}$ in $T$-dependent magneto-conductance analysis, we fitted the data to Eq. 2 with a constant Drude conductance $G_0$ and $L_\varphi \propto T^{-1/3}$. The fits are shown as the solid lines. The functional form of the fits is consistent with the data, except for some deviation at $T<10$K which is believed to be due to the saturation of $L_\varphi$ shown in Fig. 2(B). However, the size of $L_\varphi$ needed to account for $G(T)$ at $B=0$ by Eq. 2 is about two times larger than values in Fig. 2(B). This apparent contradiction leads us to conclude that, in addition to weak localization, there is another mechanism introducing a conductance correction proportional to $L_\varphi$ to the Drude conductance. Thus the conductance correction due to weak localization is responsible for about half of the conductance drop of the sample as the temperature decreases at $B=0$. This is consistent with the weak insulating-like temperature dependence of the Drude conductance obtained in the magneto-conductance analysis, as shown in Fig.2(C).

In conclusion, we studied the temperature dependence of the magneto-conductance in a single InAs nanowire. One dimensional weak localization theory describes our data well and is used to extract the phase coherence length at different temperatures and gate voltages. The temperature dependence of the coherence length shows that the main dephasing mechanism is the one-dimensional electron-electron scattering mechanism. Moreover, as the electron density is tuned by a gate voltage, we observed in the $G(B)$ data a possible signature of the suppression of weak localization in multiple 1D subbands with different mobilities. Our results illustrate the potential of chemically synthesized InAs nanowires for studying coherent electron transport with tunable 1D subband fillings.

**ACKNOWLEDGMENT** X.P.A. Gao acknowledges CWRU startup fund and ACS Petroleum Research Fund for supporting this work.

elementary electron charge, $h$ is the thickness of the oxide, $r$ is the radius of the NW, $\varepsilon$ and $\varepsilon_0$ are the dielectric constant of $SiO_2$ and permittivity of the free space respectively. Threshold voltage $V_{th} \sim -15V$.